\documentclass[final,3p,times,twocolumn]{elsarticle}
\usepackage{moreverb}
\usepackage{graphicx,wrapfig,float}
\usepackage{capt-of}
\usepackage[cmex10]{amsmath}
\usepackage{array}
\usepackage{amssymb}
\usepackage{color}
\usepackage{stfloats}
\usepackage{tabularx}
\usepackage[table]{xcolor}
\usepackage{hyperref}

\usepackage{booktabs}
\usepackage{multirow,graphicx}
\usepackage{tcolorbox}
\usepackage{colortbl}
\tcbuselibrary{skins}
\usepackage{caption}
\usepackage{amsmath}
\usepackage{natbib}

\journal{--}

\begin{document}
\begin{frontmatter}

\title{A rule-based method to model myocardial fiber orientation in cardiac biventricular geometries with outflow tracts}

\author[1]{R. Doste\corref{cor1}}
\ead{ruben.doste@upf.edu}
\cortext[cor1]{Corresponding author}
\author[1]{D. Soto-Iglesias}
\author[1]{G. Bernardino}
\author[1]{A. Alcaine}
\author[2]{R. Sebastian}
\author[3]{S. Giffard-Roisin}
\author[3]{M. Sermesant}
\author[4]{A. Berruezo}
\author[5]{D. Sanchez-Quintana}
\author[1]{O. Camara}

%\authormark{DOSTE \textsc{et al}}

\address[1]{Physense, Department of Information and Communication Technologies, Universitat Pompeu Fabra, Barcelona, Spain}

\address[2]{Computational Multiscale Simulation Lab (CoMMLab), Department of Computer
Science, Universitat de Valencia, Valencia, Spain}

\address[3]{Inria, Asclepios Research Group, Sophia-Antipolis, France}

\address[4]{Arrhythmia Section, Cardiology Department, Thorax Institute, Hospital Clinic, Universitat de Barcelona, Barcelona, Spain}

\address[5]{Department of Anatomy and Cell Biology, Faculty of Medicine, University of Extremadura, Badajoz, Spain}

%\corres{*Ruben Doste. \email{ruben.doste@upf.edu}}

%\presentaddress{Present address}

\begin{abstract}
{ Rule-based methods are often used for assigning fiber orientation to cardiac anatomical models. However, existing methods have been developed using data mostly from the left ventricle. As a consequence, fiber information obtained from rule-based methods often does not match histological data in other areas of the heart such as the right ventricle, having a negative impact in cardiac simulations beyond the left ventricle. In this work, we present a rule-based method where fiber orientation is separately modeled in each ventricle following observations from histology. This allows to create detailed fiber orientation in specific regions such as the endocardium of the right ventricle, the interventricular septum and the outflow tracts. We also carried out electrophysiological simulations involving these structures and with different fiber configurations. In particular, we built a modelling pipeline for creating patient-specific volumetric meshes of biventricular geometries, including the outflow tracts, and subsequently simulate the electrical wavefront propagation in outflow tract ventricular arrhythmias  with different origins for the ectopic focus. The resulting simulations with the proposed rule-based method showed a very good agreement with clinical parameters such as the 10 ms isochrone ratio in a cohort of nine patients suffering from this type of arrhythmia. The developed modelling pipeline confirms its potential for an in silico identification of the site of origin in outflow tract ventricular arrhythmias before clinical intervention.}
\end{abstract}
\end{frontmatter}

%\maketitle

\section{Introduction}

Personalized electrophysiological simulations have shown promising results to support clinical decisions in cardiology~\cite{Arevalo2016,Lopez-Perez2015Three-dimensionalApplications}. One important factor that affects these simulations is how myocardial fiber orientation is established in the heart model. As described in previous works~\cite{Roberts1979InfluenceDog,Punske2005EffectHearts,Chen1993EffectsFibrillation.,Waldman1988RelationVentricle.,Lunkenheimer2006TheMass,Smerup2013AMyocardium,Pluijmert2016DeterminantsOrientation}, cardiomyocyte orientation determines the preferential electrical wave propagation and tissue contraction in the heart. Therefore, a proper orientation of the myofibers (aggregations of cardiomyocytes) is needed to obtain valid and accurate simulation results. However, determination of the 3D architecture of myofibers has been a challenge for anatomists among centuries. Only in the last 30 years, due to recent advances in microscopy and medical imaging, fully detailed descriptions have been obtained~\cite{Sanchez-Quintana1990MyocardialHeart,Scollan2000ReconstructionImaging,Jouk2000Three-dimensionalHeart.,Gonzalez-Tendero2017WholeTomography,Varray2017ExtractionAnalysis,Stephenson2017HighModeling}. Imaging techniques such as diffusion tensor magnetic resonance imaging (DT-MRI), micro-CT or X-ray phase-contrast imaging, make possible to acquire information about myofiber distribution. Unfortunately, most of these imaging techniques can only be applied on ex-vivo specimens since they need long acquisition and reconstruction times to collect accurate myofiber information. Researchers are currently developing in vivo DT-MRI sequences~\cite{Toussaint2013InProcessing,Nielles-Vallespin2013InApproaches,vonDeuster2016StudyingImaging.}, showing very promising results. Yet these advanced imaging techniques cannot easily be applied to patients nowadays and are still limited to a reduced number of 2D slices followed by 3D interpolation techniques, providing coarse spatial resolution and low signal-to-noise ratios.

Due to the difficulties to acquire patient-specific data, there are two main options for incorporating myofiber orientation into 3D computational models of the heart: fitting of a map of fibers extracted from ex-vivo data; or using rule-based methods (RBM). Fitting of fiber maps can be achieved using atlas-based methods, which warp a template or an average atlas of fibers obtained by imaging techniques into a new heart geometry~\cite{Peyrat2007AHearts,Lombaert2011StatisticalDT-MRI,Lombaert2012HumanPopulation} or using statistics-based predictive techniques for assigning the fibers~\cite{Lekadir2014StatisticalPredictors}. However, these methods require complex registration algorithms to establish correspondences between different heart geometries and are highly dependent on the quality of the original data. The second option involves rule-based methods. RBMs are based on mathematical descriptions of myofiber data acquired from experimental observations and have become the most common strategy to incorporate fiber information in cardiac computational models. Since the study by Streeter~\cite{Streeter1969FiberSystole.}, where fiber orientation was studied in canine hearts, a high number of RBMs have been developed and used by the scientific community due to their relatively easy adaptation to any geometrical model~\cite{Nielsen1991MathematicalHeart.,Bayer2012,Wong2012GeneratingInterpolation.,Nagler2013PersonalizationFilterb}.

\begin{figure*}[ht!]
\centering
\includegraphics[width=0.99\textwidth]{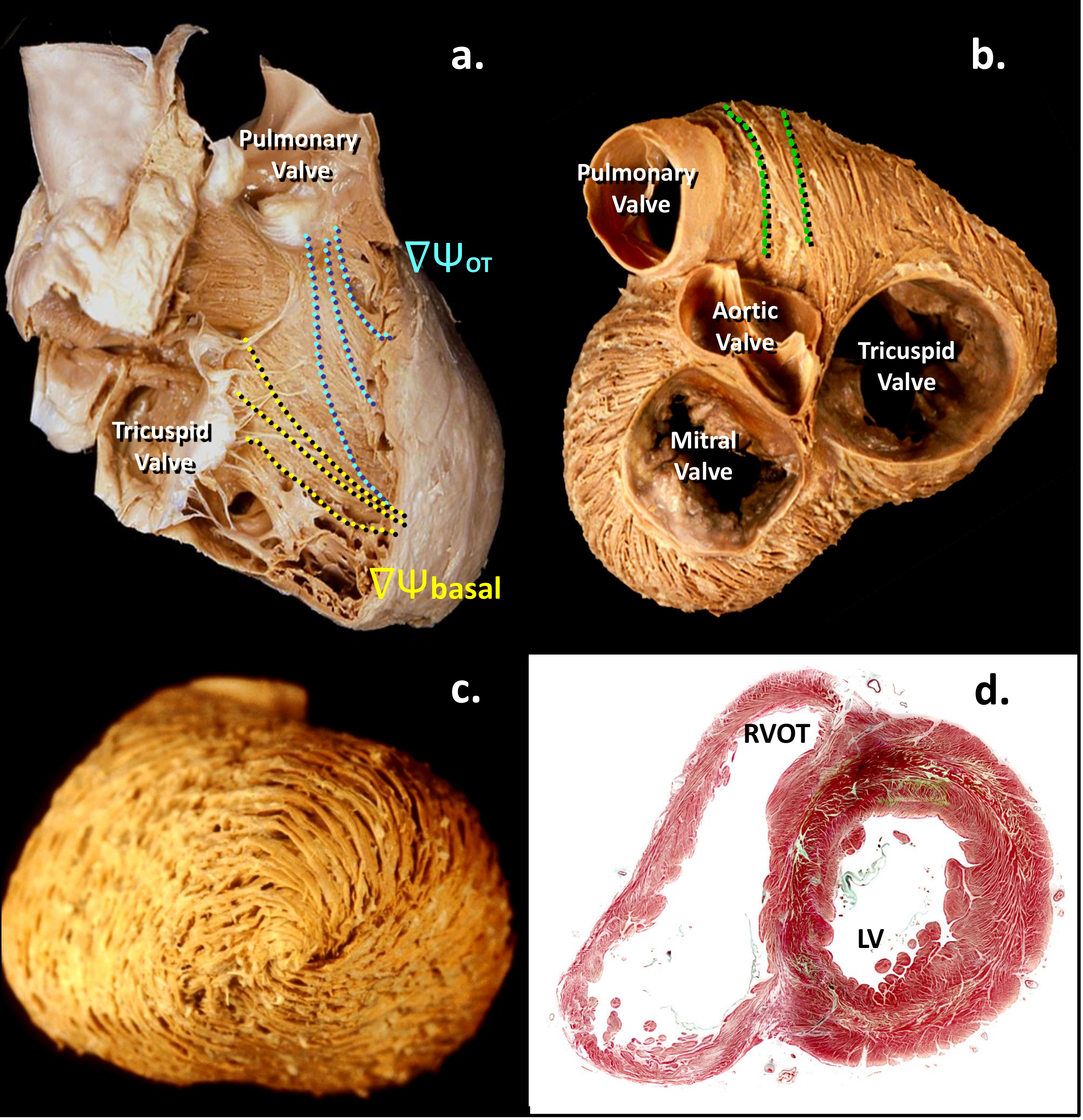}
\caption{Histological data of the heart. a) Fiber configuration in the RV sub-
endocardium, with longitudinal directions to the pulmonary and tricuspid valves
(dashed blue and yellow lines, respectively). b) Epicardial fiber configuration of both ventricles (top view). The dashed green line represent circumferential fibers of the OT. c) Apical view of the fiber epicardial layer d) Short axis slice of the heart showing transmural fiber orientation. $\nabla\Psi_{basal}$: apico-basal direction; $\nabla\Psi_{OT}$ : apico-OT direction; RVOT: Right Ventricle Outflow Tract; LV: Left Ventricle}
\label{fig: Hist}
\end{figure*}
The two described strategies have the common issue of being mainly focused on the left ventricle (LV). This is due to the complexity of obtaining accurate data of the right ventricle (RV), especially on fiber orientation. For instance, according to histological data, fiber orientation in the RV sub-endocardium has a longitudinal direction from apex towards pulmonary and tricuspid valves, as illustrated in Figure~\ref{fig: Hist}, which is not the case in LV-based RBM. The outflow tract (OT) of the ventricles is a structure that plays a key role in some pathologies like  outflow tract ventricular arrhythmia (OTVA). It has a particular fiber configuration with longitudinal and circumferential directions in the sub-endocardium and sub-epicardium, respectively. Unfortunately, most biventricular computational models include an artificial basal plane well below the valves rather than a complete biventricular geometry, thus not including the outflow tracts. 

To overcome these issues, we have developed a RBM that includes specific fiber orientation in different cardiac regions such as the RV endocardium, the interventricular septum and the outflow tracts, following observations from histological data. This outflow tract extended RBM (OT-RBM) allows running in silico simulations modelling pathologies where these regions are relevant such as OTVAs and Tetralogy of Fallot, among others. The OT-RBM processes both ventricles independently, which gives more flexibility to generate different fiber configurations. Therefore, septal fiber orientation can also be independently modified, allowing the study of its discontinuity, which is still under debate~\cite{Agger2016InsightsZone,Boettler2005NewStudy.,Kocica2006TheMyocardium}. In this work, the introduced RBM is compared with state-of-the-art fiber generation models such as the ones based on Streeter observations~\cite{Streeter1969FiberSystole.} and Bayer et al.~\cite{Bayer2012}, as well as with DT-MRI data. Additionally, electrophysiological simulations with fiber orientation provided by the OT-RBM are performed in a set of nine patient-specific OTVA biventricular geometries. Simulation results are then compared to clinical observations from electro-anatomical maps of these patients for different sites of origin of the ectopic focus.

\begin{figure*}[ht]
\centering
\includegraphics[width=0.99\textwidth]{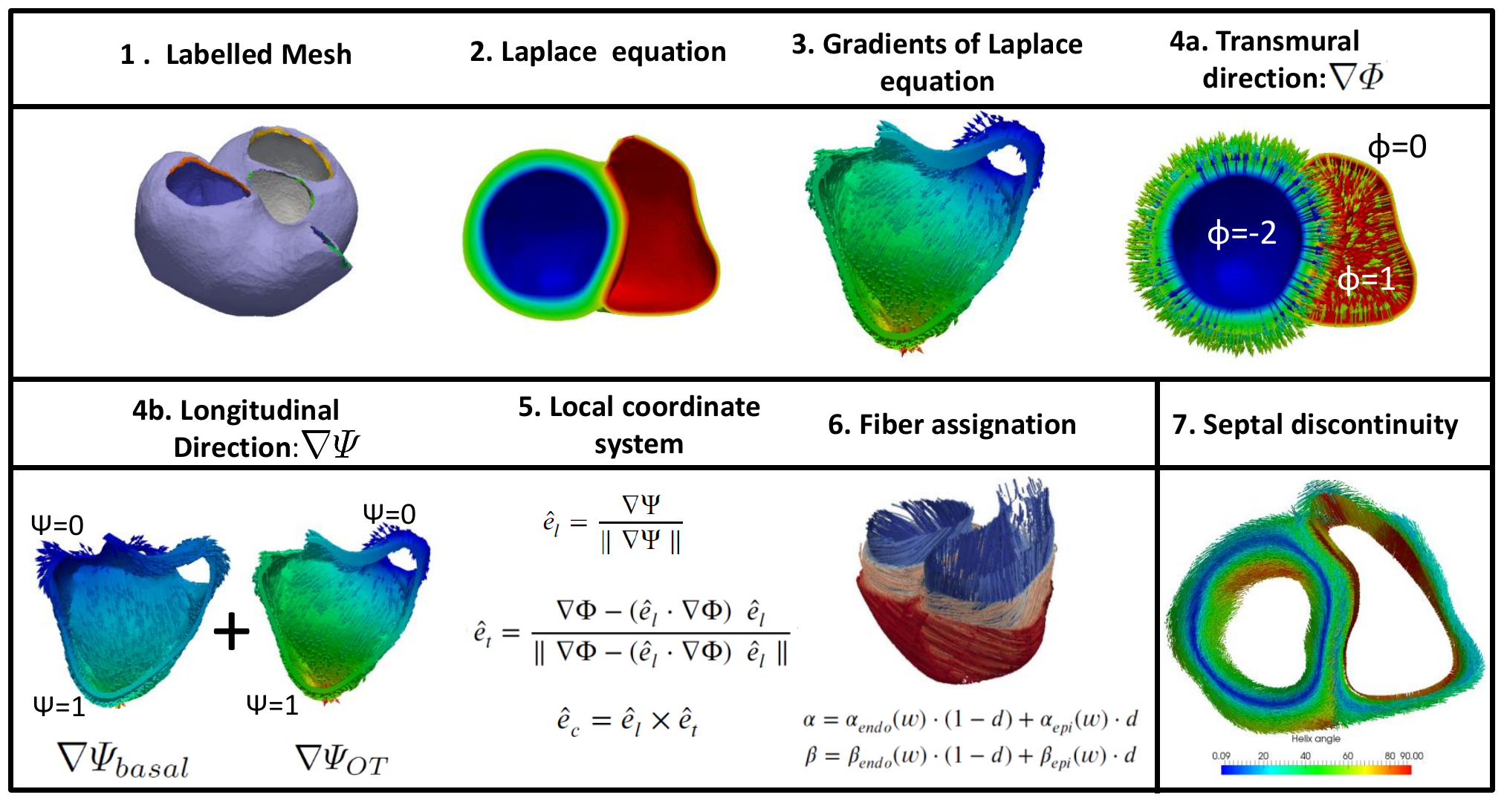}
\caption{Scheme showing the different steps of the OT-RBM. $\nabla\Psi$: longitudinal gradient; $\nabla\Psi_{basal}$: apico-basal gradient; $\nabla\Psi_{OT}$: apico-OT gradient; $\nabla\Phi$: transmural gradient; $\hat{e}_{l}$: longitudinal axis; $\hat{e}_{t}$: transmural axis; $\hat{e}_{c}$: circumferential axis. }
\label{fig: Scheme}
\end{figure*}

\section{Methods} 
%\subsection{Overview of the method}
\subsection{General overview }

The OT-RBM uses the Laplace-Dirichlet method introduced in the work of Bayer et al.~\cite{Bayer2012} (which is mainly based on LV fiber observations) to create a new local coordinate system in each point of the mesh.  This coordinate system allow assigning specific fiber information in the RV  according to histological data and extending it to the outflow tracts of both ventricles. Additionally, the proposed RBM allows, if necessary, septal fiber discontinuities between the LV and RV.
 The whole pipeline to develop the new RBM is summarized in Figure~\ref{fig: Scheme}. It starts with the generation of volumetric labelled biventricular meshes, representing cardiac structures including outflow tracts from both ventricles up to the valve planes. Subsequently, the Laplace equation is solved with different boundary conditions to define the directions (i.e. transmural and longitudinal) required for the local coordinate system that will guide fiber assignment at each point of the mesh. The longitudinal direction in both ventricles is defined as a combination of the vector fields resulting from the Laplace equation between the apex and the two valves. Once the directions are computed, a local coordinate system is generated at each point of the mesh. Fiber orientation is finally estimated for each cardiac structure after finding the most appropriate angles of the coordinate system to match histological observations.

\subsection{Volumetric labelled mesh generation}
Biventricular geometries used in this work were represented by patient-specific tetrahedral meshes built from the processing of computed tomography (CT) images. These images corresponded to nine patients with idiopathic OTVAs submitted for ablation procedure at Hospital Cl\'inic de Barcelona. A MDCT ECG-gated study was performed on a 128 x 2 -slice CT scanner (Somatom Definition Flash, Siemens Healthcare, Erlangen, Germany). Images were acquired during an inspiratory breath-hold using retrospective ECG-gating technique with tube current modulation set between 50\% and 100\% of the cardiac cycle. The isotropic spatial resolution was 0.4 x 0.4 x 0.4 mm.

The biventricular geometries, including the outflow tracts and valve planes, were obtained from CT images using a semi-automatic segmentation procedure with region growing techniques available in the 3DSlicer~\footnote{\url{https://www.slicer.org}} software. Subsequently, surface meshes were generated from the obtained binary segmentations applying the classical Marching Cubes method, which was followed by some post-processing steps (e.g. smoothing, labelling) performed in Blender~\footnote{\url{https://www.blender.org}}. Finally, tetrahedral meshes ($\sim$80000 nodes and $\sim$400000 elements) were created using the iso2mesh~\footnote{\url{http://iso2mesh.sourceforge.net}} tool. 

The next step of the methodology was mesh labelling, where different cardiac geometrical surfaces were identified to apply the Dirichlet conditions and construct the local reference system. These surfaces were the epicardium, the RV and LV endocardial walls, RV and LV apices and the four heart valves. These regions were easy to identify in the studied CT images, enabling the application of the OT-RBM in a wide variety of heart geometries. Mesh labelling is usually performed either manually, by tagging surfaces during the segmentation process~\cite{Crozier2016Image-BasedModeling}, or using semi-automatic techniques~\cite{Bayer2012,A.PalitG.A.TurleyS.K.Bhudia2014AssigningModel}. We used a semi-automatic approach where the apices and valve position were manually selected in CT images and endocardial and epicardial surfaces were automatically identified using ray/triangle intersection algorithms~\cite{Moller1997FastIntersection}. Rays were traced from the normal of every face of the surface mesh and intersections with other faces were computed. If intersections were found, the face was classified as endocardium; the rest were set as epicardium. Figure~\ref{fig: Scheme} (Step 1) shows the resulting mesh labelling for one of the processed biventricular geometries. Each mesh region is identified by a different color.

\subsection{Local coordinate system}
Cardiac fibers in finite-element meshes are represented in each node by an unitary 3D vector that is oriented in the preferential direction of the electrical propagation and mechanical contraction of the myofiber. Therefore, a local orthonormal coordinate system is needed to define the myofiber vector in all mesh nodes. The orthotropic axes of the reference system are the longitudinal ($\hat{e}_{l}$), transmural ($\hat{e}_{t}$) and local circumferential ($\hat{e}_{c}$) directions. Transmural and longitudinal directions can be defined by solving the Laplace equation using the corresponding surfaces as Dirichlet boundary conditions and computing the gradient of the solution, as in Bayer et al.~\cite{Bayer2012}. In the OT-RBM these geometrical surfaces were the RV and LV endocardial walls, the whole bi-ventricular epicardium, the RV and LV apices and the four cardiac valves (the tricuspid and pulmonary valves for the RV and the mitral and aortic valves for the LV). The local circumferential direction was defined as the cross product of transmural and longitudinal directions. Fiber orientation was then obtained by rotating the obtained vector $\hat{e}_{c}$  by a given angle $\alpha$ and $\beta$ to match histological observations. 

\begin{figure}[ht]
\centering
\includegraphics[width=0.99\columnwidth]{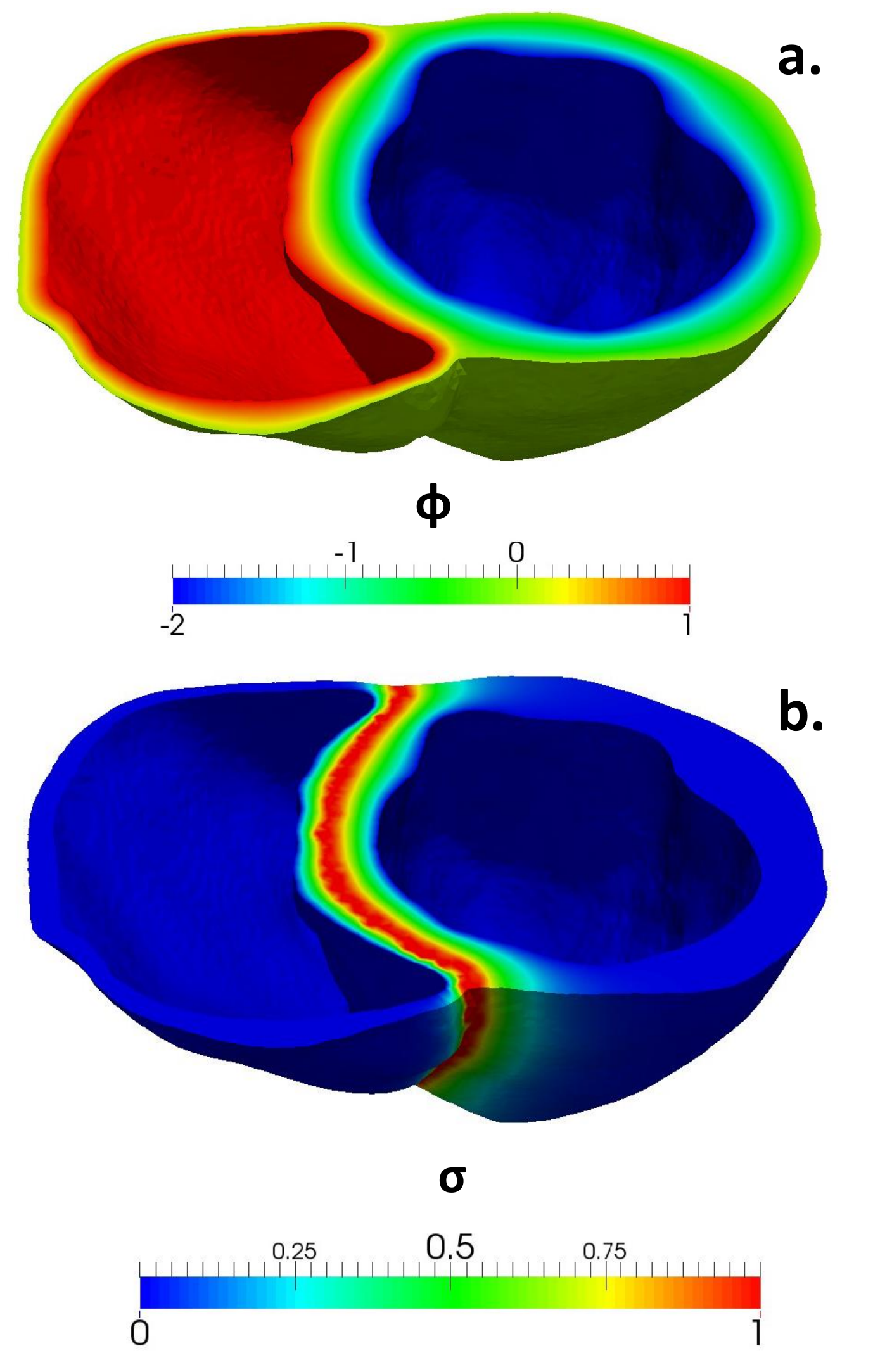}
\caption{a) Example of a transmural map obtained by solving Laplace equation. b) Solution of solving the Laplace equation between the inter-ventricular septal surface and the RV and LV endocardium}
\label{fig: Laplace}
\end{figure}

Transmural direction ($\nabla\Phi$) was obtained by solving the Laplace equation between the endocardium of each ventricle and the epicardium; subsequently the gradient of the Laplace solution (Figure~\ref{fig: Laplace}a) was computed. Negative and positive values were correspondingly assigned to the LV and RV endocardium ($\Phi=-2$ and $\Phi=1$, respectively) as Dirichlet boundary conditions. The epicardium was assigned to a zero value ($\Phi=0$), thus allowing a discrimination between the two ventricles: positive and negative values were assigned to the RV and LV, respectively. These boundary conditions allowed to replicate findings from  histological studies that affirm that two thirds of the septum belong to the LV and one third to the RV~\cite{Gonzalez-Tendero2017WholeTomography,Boettler2005NewStudy.,Sanchez-Quintana2015AnatomicalElectrophysiologist,Agger2016TheModel}.  Laplace equations and its gradients were solved using Elmer software~\footnote{\url{https://www.csc.fi/web/elmer}}.

The longitudinal direction ($\nabla\Psi$) was defined separately for each ventricle. This direction was the result of the weighted sum of the apico-basal ($\nabla\Psi_{basal}$) and apico-OT ($\nabla\Psi_{OT}$) gradients, defined individually in both ventricles: $\nabla\Psi_{basal}$ considered the apex-mitral valve and the apex-tricuspid valve directions in the LV and the RV, respectively; $\nabla\Psi_{OT}$ followed the apex-aortic valve and the apex-pulmonary valve directions in the LV and RV, respectively. These directions were already described by Greenbaum et al.~\cite{Greenbaum1981} and can be visualized in Figure~\ref{fig: Hist} (dashed lines). The resulting longitudinal fiber direction for each ventricle was set as follows:

\begin{equation} 
\nabla\Psi = \nabla\Psi_{basal}\cdot w + \nabla\Psi_{OT}\cdot(1-w) 
\end{equation}

\begin{figure}[ht]
\centering
\includegraphics[width=0.99\columnwidth]{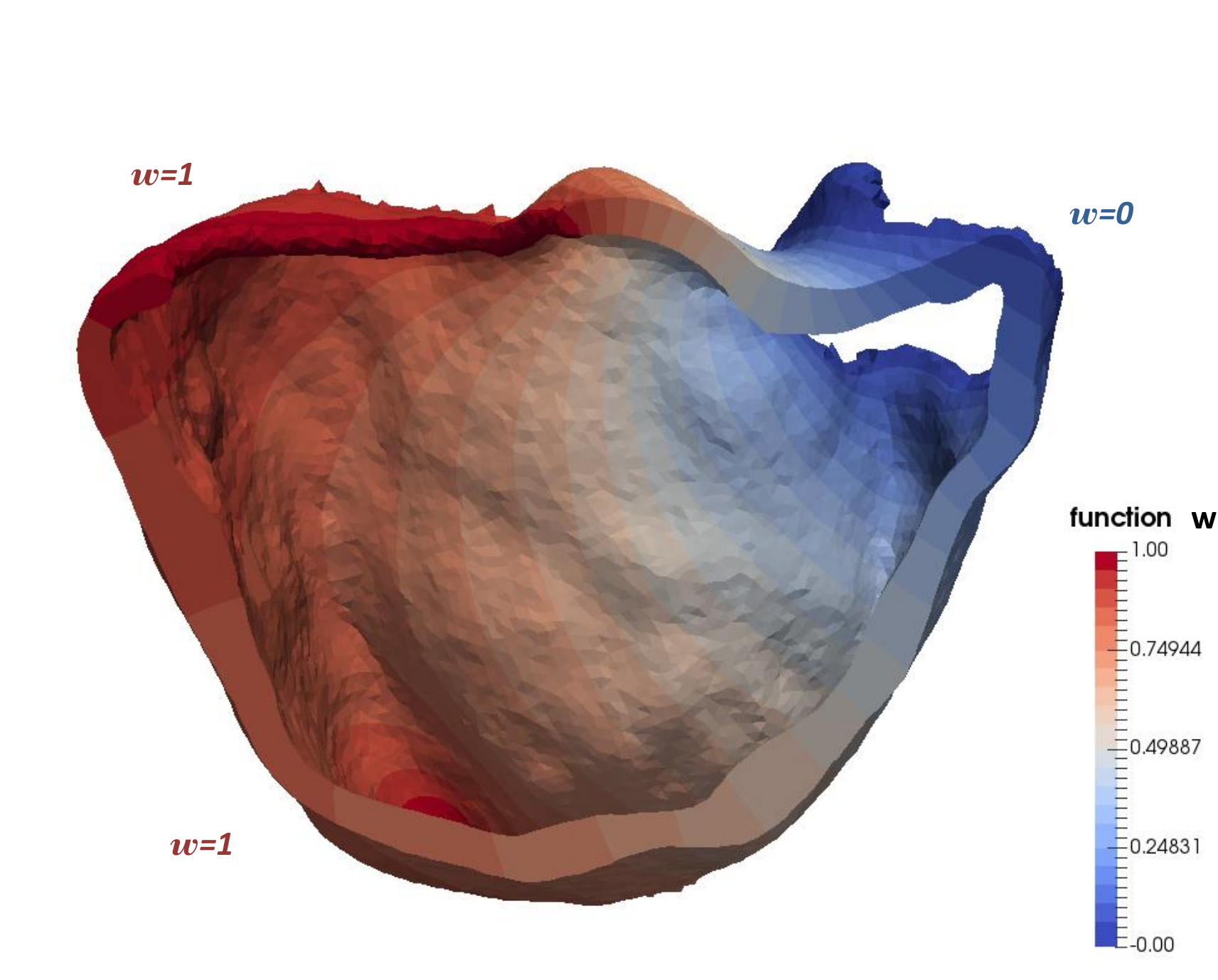}
\caption{Example of intra-ventricular interpolation function $w$ in the RV. The obtained values will guide the fiber interpolation between the apico-basal and apico-OT longitudinal directions within the ventricle.}
\label{fig: Functionf}
\end{figure}

The sum of the apico-basal and apico-OT gradients was weighted by an intra-ventricular interpolation function $w$, which was computed in the RV by solving the Laplace equation between the apex ($w=1$), tricuspid valve ($w=1$) and pulmonary valve ($w=0$), as shown in Figure~\ref{fig: Functionf}. In the LV, the equation was solved involving the apex ($w=1$), the mitral valve ($w=1$) and the aortic valve ($w=0$). In this way, we obtained a smooth distribution of values following the cardiac surface allowing to properly define the OTs and control the smoothness in fiber changes near the OT in different geometries.

Using the previously calculated gradients ($\nabla\Psi$,$\nabla\Phi$), the local coordinate system was set up for each mesh node, which is fully described with the following vectors: 
\begin{align} 
&\hat{e}_{l} = \frac{\nabla\Psi}{\parallel\nabla\Psi\parallel}   
&\hat{e}_{t} = \frac{\nabla\Phi-(\hat{e}_{l}\cdot\nabla\Phi)\enskip\hat{e}_{l}}{\parallel\nabla\Phi-(\hat{e}_{l}\cdot\nabla\Phi)\enskip\hat{e}_{l}\parallel} \quad
&\hat{e}_{c} = \hat{e}_{l} \times \hat{e}_{t} 
\end{align}

\subsubsection{Rotation of local coordinate system to match histological observations}
  Fiber orientation was finally obtained in every mesh node by rotating the local coordinate system to better match histological observations, as follows: 1)   Vector $\hat{e}_{c}$ was rotated counterclockwise around $\hat{e}_{t}$  by an angle $\alpha$; and 2) subsequently, vector $\hat{e}_{c}$  was again rotated counterclockwise around $\hat{e}_{l}$ by a transverse angle $\beta$ . Note that the definition of angle $\alpha$ (angle with respect to the local circumferential direction) is not equivalent to the $\alpha$ helix angle defined in other histological studies or RBMs. The main reason is that our longitudinal direction is not always oriented in the apico-basal direction since it also has a component pointing in the apex-OT direction, which is the predominant one in the OTs. Far from the OTs, where the apex-OT component is negligible, the longitudinal direction only points towards the atrioventricular valves and therefore, our angle  $\alpha$  and the helix angle match. 

The first rotation was defined in each ventricle as a counterclockwise rotation of the vector $\hat{e}_{c}$ around $\hat{e}_{t}$ with an angle $\alpha$:
\begin{flalign}
\alpha = \alpha_{endo}(w) \cdot (1-d) + \alpha_{epi}(w)\cdot d
\end{flalign}
\noindent where $d$ is the transmural depth normalized from 0 to 1. The different values of $\alpha_{endo}$ and $\alpha_{epi}$ were chosen to replicate the following observations from several histological studies~\cite{Sanchez-Quintana1990MyocardialHeart,Sanchez-Quintana2015AnatomicalElectrophysiologist,Greenbaum1981,Ho2006AnatomyDimensions.}:
\begin{itemize}
\item Left ventricle (based on Greenbaum's observations ~\cite{Greenbaum1981}): $\alpha_{endo}(w=1)=-60^{\circ}$; $\alpha_{epi}(w=1)=60^{\circ}$
\item Right ventricle (based on Greenbaum~\cite{Greenbaum1981}, Ho~\cite{Ho2006AnatomyDimensions.} and Sanchez-Quintana~\cite{Sanchez-Quintana2015AnatomicalElectrophysiologist}): $\alpha_{endo}(w=1)=90^{\circ}$ (same as longitudinal direction); $\alpha_{epi}(w=1)=-25^{\circ}$
\item Outflow tracts (based on Sanchez-Quintana~\cite{Sanchez-Quintana2015AnatomicalElectrophysiologist}): $\alpha_{epi}(w=0)=0^{\circ}$ (circumferential direction); $\alpha_{endo}(w=0)=90^{\circ}$ (longitudinal direction)
\end{itemize}

The expression for the transverse angle, $\beta$, was the following:
\begin{flalign}
\beta = \beta_{endo}(w) \cdot (1-d) + \beta_{epi}(w)\cdot d
\end{flalign}

\noindent Values of the transverse angle were derived from several studies~\cite{Greenbaum1981,Stephenson, Lunkenheimer2013ModelsCardiologists} as follows:
\begin{itemize}
\item Left ventricle:

$\beta_{endo}(w=1)=-20^{\circ}$  and    $\beta_{epi}(w=1)=20^{\circ}$
\item Right ventricle:

$\beta_{endo}(w=1)=0^{\circ}$   and  $\beta_{epi}(w=1)= 20^{\circ}$
\item Outflow tracts:

$\beta_{endo}(w=0)=0^{\circ}$ and   $\beta_{epi}(w=0)=0^{\circ}$
\end{itemize}

\subsubsection{Septal configuration}
The OT-RBM allowed to control fiber angles in the septum, since we previously divided our septal geometry into RV and LV using the transmural direction. Therefore, the intersection surface between both ventricles (i.e. inter-ventricular septal surface) can be used to guide the interpolation of fiber angles in the septum. This is done by solving the Laplace equation between the inter-ventricular septal surface and the RV and LV endocardium (Figure~\ref{fig: Laplace}, b). The obtained values ($\sigma$) were used for forcing a smooth transition in both ventricles from the initial fibers in the septal surface ($\alpha_{septal}$). Hence, the final expression for assigning the fiber angle in the septum, which could easily be modified to enforce continuity or a certain angle of discontinuity, was the following:
\begin{flalign}
\alpha_{final} = \alpha \cdot (1-\sigma) + \alpha_{septal}\cdot \sigma
\end{flalign}

\begin{figure*}[ht]
\centering
\includegraphics[width=0.99\textwidth]{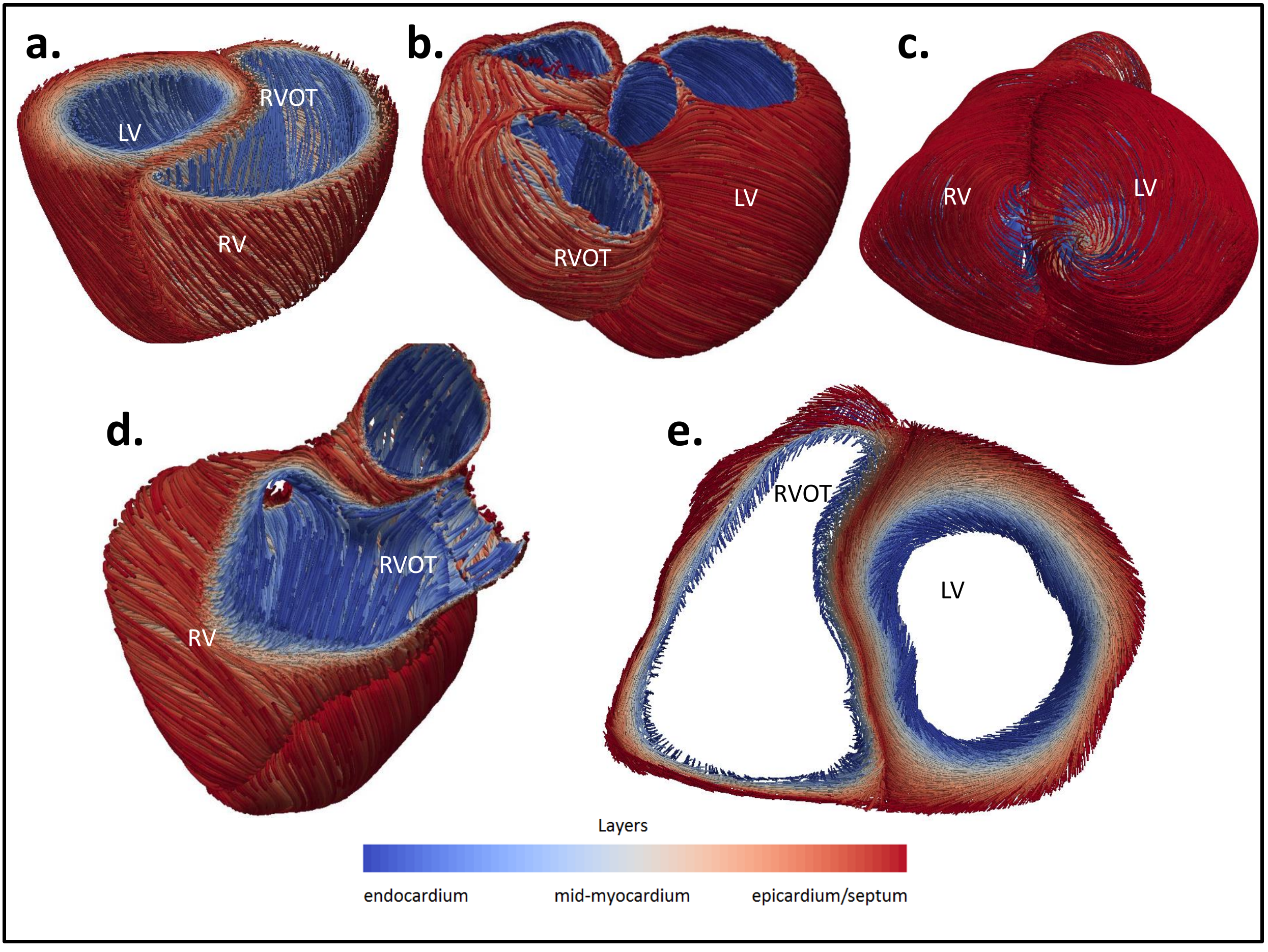}
\caption{Fiber orientation obtained with the OT-RBM for different biventricular geometries and regions of the heart. LV: left ventricle; RV: right ventricle; RVOT: right ventricular outflow tract.}
\label{fig: Geometries2}
\end{figure*}

\section{Experimental results}
Results of applying the developed method to different geometries can be seen in  Figure~\ref{fig: Geometries2}. They show a good agreement with histological data presented in Figure~\ref{fig: Hist}, especially in certain areas: right ventricle outflow tract (RVOT) fiber orientation (Figure~\ref{fig: Hist}a and Figure~\ref{fig: Geometries2}a, b, d); epicardial fiber configuration (Figure~\ref{fig: Hist}b, c and Figure~\ref{fig: Geometries2}b, c); and the transmural variation in LV fibers (Figure~\ref{fig: Hist}d and Figure~\ref{fig: Geometries2}e). A more exhaustive and quantitative evaluation of OT-RBM performance was conducted by means of three experiments, described below. 
\subsection{Experiments}
Three experiments were designed and conducted to analyze the performance of the proposed RBM. In the first experiment we estimated the differences between fiber configurations provided by the OT-RBM and existing ones in the literature. Magnetic resonance imaging data from an ex-vivo human heart available at the Johns Hopkins database~\footnote{http://cvrgrid.org/data/ex-vivo}, which also includes fiber information from DT-MRI was used for this experiment. Differences were computed as the angle between fibers in each point of the mesh. Furthermore, mean angle differences were evaluated in different regions of the heart, dividing the LV into the 17 AHA regions according to the well-known description of Cerqueira et al.~\cite{Cerqueira2002StandardizedHeart}. For the RV, a division in 15 regions based on the model proposed by Zhong et al.~\cite{Zhong2012RightFallot} was performed.
\begin{table*}[t]
\centering
\caption{ Clinical ratio (Longitudinal/Perpendicular diameter) of the 10 ms isochrones map area obtained from the simulated OTVAs using different SOO and RBMs. (RVOT: right ventricle outflow tract; RCC: right coronary cusp; LCC: left coronary cusp; OT-RBM: Outflow Tract extended rule based method; ST-RBM: Streeter-Based rule based method)}
\label{Table}
\resizebox{\textwidth}{!}{\begin{tabular}{cccccccccccc}
\multicolumn{12}{c}{\textbf{Patient}}                                                                                                                                                                                                                                                                                \\ \hline
\multicolumn{1}{c|}{}                                                                                        & \multicolumn{1}{c|}{}              & \textbf{P1} & \textbf{P2} & \textbf{P3} & \textbf{P4} & \textbf{P5} & \textbf{P6} & \textbf{P7} & \textbf{P8} & \multicolumn{1}{c|}{\textbf{P9}} & \textbf{All}  \\ \hline
\multicolumn{1}{c|}{\textbf{}}                                                                               & \multicolumn{1}{c|}{\textbf{RVOT}} & 2.06        & 1.59        & 2.07        & 2.15        & 1.81        & 1.70        & 1.78        & 1.99        & \multicolumn{1}{c|}{1.55}        & 1.86$\pm$0.22 \\
\multicolumn{1}{c|}{\textbf{OT-RBM}}                                                                        & \multicolumn{1}{c|}{\textbf{LVOT-RCC}}  & 1.08        & 0.99        & 0.62        & 1.14        & 1.00        & 1.23        & 0.99        & 1.24        & \multicolumn{1}{c|}{0.84}        & 1.02$\pm$0.2  \\
\multicolumn{1}{c|}{\textbf{}}                                                                               & \multicolumn{1}{c|}{\textbf{LVOT-LCC}}  & 0.46        & 0.41        & 0.73        & 0.56        & 0.76        & 0.95        & 0.79        & 0.78        & \multicolumn{1}{c|}{0.75}        & 0.74$\pm$0.25 \\ \hline
\multicolumn{1}{c|}{\multirow{3}{*}{\textbf{\begin{tabular}[c]{@{}c@{}}ST-RBM\end{tabular}}}} & \multicolumn{1}{c|}{\textbf{RVOT}} & 0.50        & 0.43        & 0.50        & 0.96        & 0.58        & 0.59        & 0.49        & 0.50        & \multicolumn{1}{c|}{0.56}        & 0.57$\pm$0.16 \\
\multicolumn{1}{c|}{}                                                                                        & \multicolumn{1}{c|}{\textbf{LVOT-RCC}}  & 0.87        & 0.77        & 0.42        & 0.95        & 1.70        & 1.31        & 1.14        & 1.35        & \multicolumn{1}{c|}{1.08}        & 1.07$\pm$0.37 \\
\multicolumn{1}{c|}{}                                                                                        & \multicolumn{1}{c|}{\textbf{LVOT-LCC}}  & 0.70        & 0.51        & 0.46        & 0.44        & 1.13        & 1.46        & 1.01        & 0.55        & \multicolumn{1}{c|}{0.39}        & 0.74$\pm$0.4  \\ \hline
\multicolumn{1}{c|}{\textbf{Clinical}}                                                                       & \multicolumn{1}{c|}{\textbf{}}     & 0.54 (RCC)  & 0.81 (LCC)  & 0.73  (RCC) & 0.67 (LCC)  & 0.82 (RCC)  & 0.64 (RCC)  & 1.63 (RVOT) & 0.71 (RCC)  & \multicolumn{1}{c|}{2.27 (RVOT)} &               \\ \hline
\end{tabular}}
\end{table*}

 Since the previous geometry did not include detailed fiber information in the OTs, we performed a second experiment where OT fibers are indirectly compared. This experiment involved running electrophysiological simulations, including fiber orientation patterns computed with the OT-RBM and the Streeter-based one, on heart geometries from nine idiopathic OTVA patients. The simulated electrical propagation waves were then compared with clinical data from the OTVA patients, acquired during the ablation procedure at Hospital Cl\'inic de Barcelona, Spain. Clinical data consisted of electro-anatomical maps acquired by the CARTO 3 system (Biosense Webster, Inc., Diamond Bar, CA, USA). Early activation sites were manually identified by clinicians during the intervention. Characteristics of the isochrones (area, axis ratio) around the earliest activated point in the RV endocardium were also manually measured by clinicians after the intervention. Isochronal characteristics provide useful information to predict the site of origin (SOO) of the ectopic focus~\cite{Acosta2015ImpactArrhythmias,Herczku2012MappingActivation.}: if the longitudinal axis of the isochrones follows the apico-basal axis, the SOO should be in the RVOT (following the fibers in the OT); while left ventricle outflow tract (LVOT) origins create more isotropic isochrones or with a larger perpendicular axis. 

Simulations were performed with the SOFA software~\footnote{http://www.sofa-framework.org} using the Mitchell-Schaeffer model~\cite{Mitchell2003AMembrane.} in patient-specific biventricular geometries. The SOO was placed in different parts of the OTs, including the one found by the clinicians during the intervention. Anisotropy conductivity ratio was set 2.5 in order to produce a higher conduction velocity along the fiber direction than in the transverse plane. The isochronal area, i.e. the surface occupied by mesh elements in the RV endocardium that has been activated at a given time (10 ms in our case) after the earliest activated point, and the clinical ratio of the simulated isochrones were calculated in each case following the same protocol than clinicians when analyzing isochrones from the electro-anatomical data~\cite{Acosta2015ImpactArrhythmias,Herczku2012MappingActivation.}. The protocol implemented by the clinicians measures the longitudinal diameter of the isochronal map area, which was defined by a line parallel to the septal projection of the RVOT longitudinal axis (perpendicular to the plane of the pulmonary valve). The defined longitudinal axis specified the perpendicular axis of the early activated area (perpendicular direction of the longitudinal axis). 

Finally, in order to demonstrate the influence of fiber orientation in  electrophysiological simulations we carried out a sensitivity analysis by changing RVOT fiber orientation by different angles and performing simulations from two different SOO (RVOT and right coronary cusp). Resulting isochrones were evaluated and compared with clinical observations.

\begin{figure*}[ht!]
\centering

\includegraphics[width=0.99\textwidth]{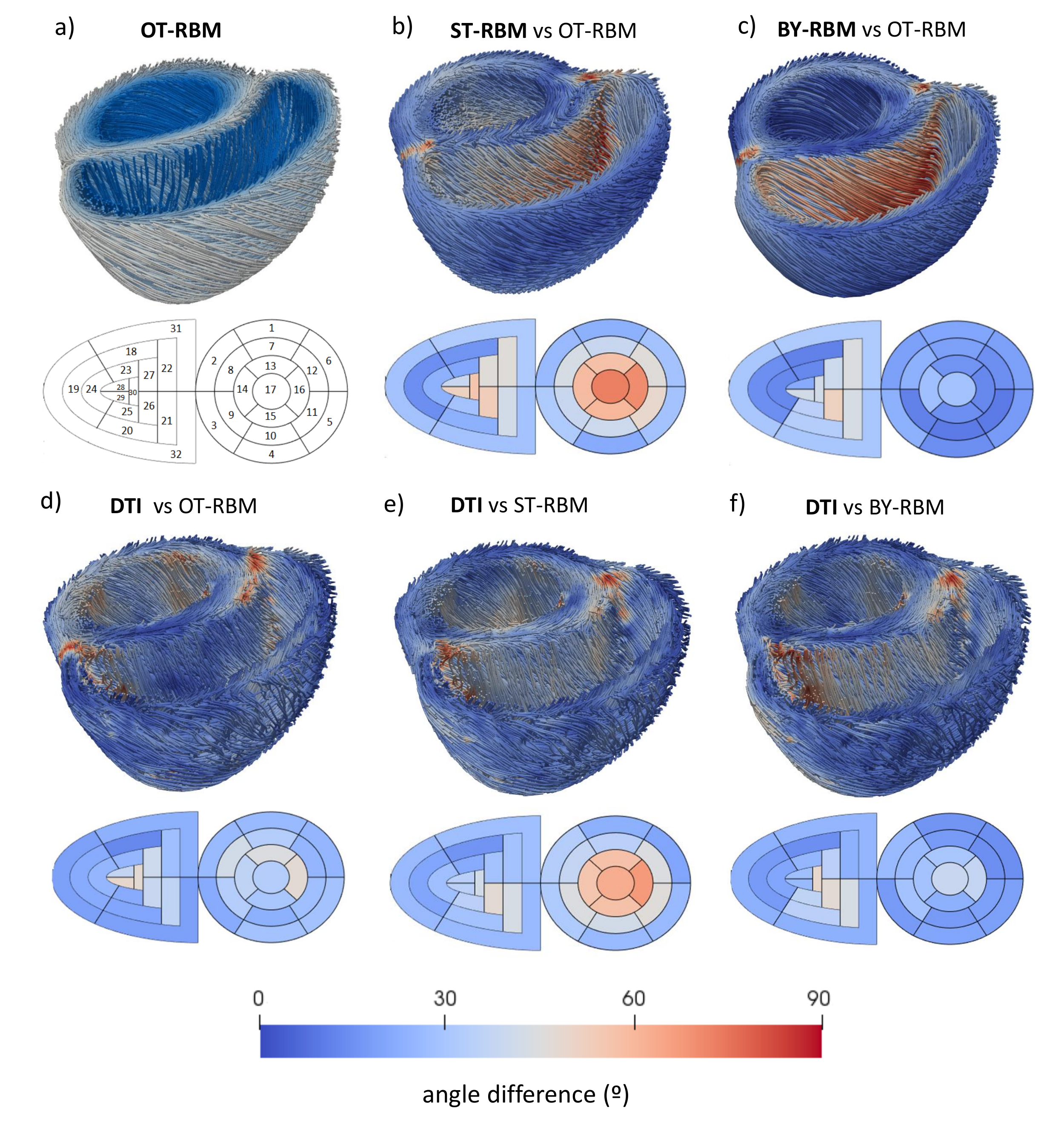}
\caption{ Top row: Fibers obtained using the OT-RBM (a) and angle differences (in degrees) with the ones obtained using  ST-RBM (b) and BY-RBM (c). Bottom row: Angle differences between  DT-MRI human fibers and OT-RBM (d), ST-RBM (e) and BY-RBM (f). Angle differences are shown in each case in a 3D mesh view and in a AHA plot with regional mean angle differences. For each geometry, name in bold indicates the represented fibers. } 
\label{fig: Diff}
\end{figure*}
\subsection{Comparison with other RBM and DT-MRI}
Figure~\ref{fig: Diff} shows the comparison of fiber distribution maps provided by the OT-RBM, the one based on Streeter (ST-RBM)~\cite{Sebastian2009CardiacPipeline}, Bayer et al.~\cite{Bayer2012} (BY-RBM) and DT-MRI on the ex-vivo human heart from the Johns Hopkins database. The top row of the figure shows the fiber configurations provided by the OT-RBM, ST-RBM and BY-RBM models (a, b, c, respectively). 
The bottom row shows the comparison between DT-MRI fibers and the different RBMs: OT-RBM (d), ST-RBM (e) and BY-RBM (f).
Information about the mean angle difference projected into a 2D visualization of different regions of the heart is also included in the figure. A quantitative analysis of these regional average angle differences is summarized in APPENDIX A. As expected, the main angle differences (red colors in Figure~\ref{fig: Diff}) between our model and the other RBMs are found in the RV endocardium, where we defined longitudinal directions from the apex towards the pulmonary and tricuspid valves. Fibers are more circumferential when generated with the other RBMs, when compared to histological data in this region. Low differences were found in the LV, especially comparing to BY-RBM (\textless 20$^\circ$). The comparison with BY-RBM also presented low differences in the free wall of the RV (\textless 15$^\circ$). Differences between OT-RBM and DT-MRI fibers are slightly larger in the LV ($\sim$30$^\circ$) but there is only one region, close to the apex, with high angle differences (\textgreater 45$^\circ$, region number 16). The ST-RBM and BY-RBM methods, especially the former, also present difficulties for generating fibers close to DT-MRI data in the LV apex (\textgreater 60$^\circ$ and $\sim$38$^\circ$ for ST-RBM and BY-RBM, respectively). 

\subsection{Electrophysiological simulations on OTVA biventricular geometries}

%\subsubsection{Isochrone comparison}

Isochrone ratios and areas were estimated for the electrophysiological simulations with RVOT and LVOT sites of origin using fiber directions computed with the OT-RBM and the ST-RBM. A quantitative analysis of isochrone ratios provided by electrophysiological simulations with the OT-RBM and the ST-RBM with different SOO in nine different patient-specific biventricular geometries is shown in Table~\ref{Table}. It also includes the SOO identified by the clinician and the manually measured isochrone ratio during the intervention. For each origin, mean of isochrone ratios were the following (mean $\pm$ SD): $1.86 \pm 0.22$ and $0.57 \pm 0.16$ for RVOT-SOO using OT-RBM and  ST-RBM, respectively; and 0.88 $\pm$ 0.3 and 0.9 $\pm$ 0.4 for LVOT-SOO, using OT-RBM and  ST-RBM respectively. We can observe than the largest differences correspond to RVOT-SOO, where the OT-RBM successfully predicted a larger longitudinal direction than the transversal one, unlike ST-RBM.

The obtained isochrone areas were the following (mean $\pm$ SD): 1.5 $\pm$ 0.4 cm$^2$ and 1.4 $\pm$ 0.4 cm$^2$ for RVOT-SOO using  OT-RBM and ST-RBM, respectively; and 5.5 $\pm$ 2.3 cm$^2$ and 5.6 $\pm$ 2.1 cm$^2$ for LVOT-SOO using  OT-RBM and  ST-RBM, respectively. Therefore, both RBM provided similar results in terms of isochrone areas, independently of the site of origin.
 \begin{figure*}{}
\centering
\includegraphics[width=0.99\textwidth]{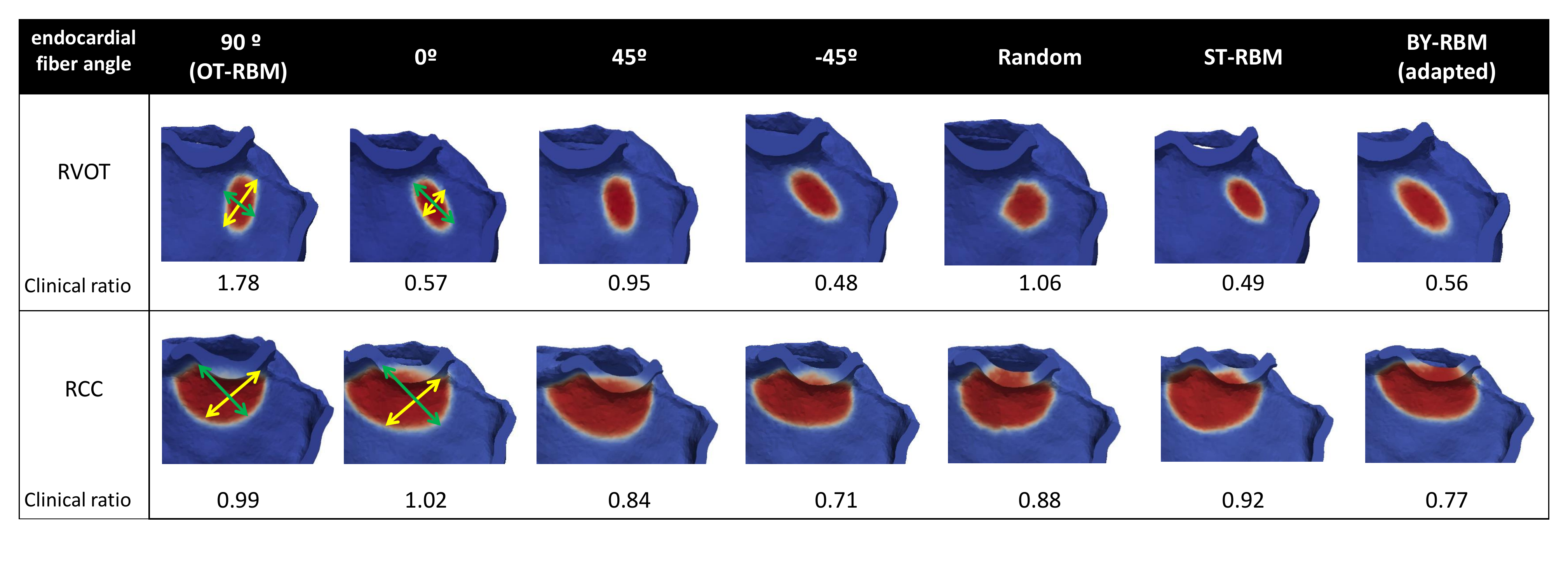}
\caption{Sensitivity analysis results for the simulated right ventricular outflow tract (RVOT) and Right coronary cusp (RCC) origin isochrones and their corresponding ratios obtained by changing endocardial fiber orientation and running electrophysiological simulations.Yellow arrows indicate the measured longitudinal direction and green arrows the perpendicular one.}
\label{fig: Sensitivity}
\end{figure*}

We also evaluated the potential of the OT-RBM as a tool to predict the SOO in OTVAs. An example of differences between isochrones obtained in the RV endocardium around the earliest activated point from the electrophysiological simulations with different SOO can be seen in the first column of Figure~\ref{fig: Sensitivity}. For LVOT origin we generated simulations from the right and left coronary cusps (RCC and LCC, respectively). Substantial differences were found in the isochrone characteristics depending on the site of origin, mainly a higher degree of isochronal isotropy with a LVOT origin (RCC in the figure) than with a RVOT origin, where the longitudinal direction along the fibers is preferential. 

We can observe that the OT-RBM contributes to simulate isochrone ratios well differentiated depending on the RV or LV site of origin. A dependent t-test between results in statistically significant differences between isochrone ratios from RVOT-SOO and RCC-SOO ($p < 0.001$) and from RVOT-SOO and LCC-SOO ($p < 0.001$), but not from RCC-SOO and LCC-SOO. Additionally, the OT-RBM successfully replicated clinical findings such as the preferential longitudinal direction in RVOT origins (isochrone ratios larger than 1.5 in P7 and P9), unlike the Streeter-based one (isochrone ratios around 0.6 in P7 and P9).

\subsection{Sensitivity analysis}

 A sensitivity analysis was carried out to assess the influence of RVOT fiber orientation in the simulated RV isochrones, with different ectopic focus in the RVOT and LVOT (RCC origin). Fibers were changed as follows: 90$^\circ$ (chosen value in the OT-RBM), 0$^\circ$, 45$^\circ$ and -45$^\circ$. We also included fiber configurations provided randomly, with ST-RBM and BY-RBM; we extended ST-RBM and BY-RBM to work in heart geometries with OT, placing the basal plane boundary condition in the mitral and tricuspid valves. Figure~\ref{fig: Sensitivity} shows the simulated 10 ms isochrones in the RV and their corresponding clinical isochrone ratio, obtained with the different fiber configurations. Longitudinal and perpendicular ratios are displayed in yellow and green arrows, respectively. Only simulations guided by OT-RBM generated fibers were able to replicate the high clinical ratio (1.78) and correct isochronal orientation obtained with a RVOT site of origin. The simulated isochrones with a LVOT-RCC origin did not present large differences with respect to the chosen fiber configuration.
 
\section{Discussion}
The aim of this work was to develop a novel rule-based method (OT-RBM) for assigning myofiber orientation in heart geometries including specific information of the RV and OTs. This new method will allow performing in silico studies to better understand and predict the behaviour of cardiac pathologies where the RV and OTs have an important role. The developed OT-RBM is based on myocyte aggregate configuration observed on histological data. It includes the following novel features that were not available before: 1) treatment of each ventricle separately, which improves the individual fitting of fiber angles; 2) new definition of longitudinal direction ($\nabla\Psi$) based on histological descriptions {\cite{Greenbaum1981,Ho2006AnatomyDimensions.}}, which generate a new coordinate system allowing simple angle assignation to different regions of the heart; 3) fiber angles specific to the LV, RV, septum and OTs; and 4) addition of septal interpolation between the ventricles. These novelties make the OT-RBM flexible enough to assimilate fiber information from multimodal imaging (DT-MRI, histology, synchrotron data) and to work in a wide variety of heart geometries.

The proposed OT-RBM was compared with state-of-the-art RBMs, namely Streeter-based RBM (ST-RBM) and Bayer's RBM (BY-RBM). Differences on fiber configuration provided by the different RBMs can be visualized in Figure~\ref{fig: Diff}. A quantitative regional comparison is also provided in Appendix A. As expected, the largest differences appear mainly near the RVOT, the RV endocardium and closer to the septum. For instance, ST-RBM and BY-RBM fibers are more circumferential that when generated with the OT-RBM, when compared to histological data in the RV. Interestingly, there were small differences (\textless 20$^\circ$) observed between OT-RBM and BY-RBM fiber configurations in the LV. This suggests that longitudinal directions obtained with the apex-mitral and apex-aortic valve boundary conditions in OT-RBM are equivalent to the apex-base ones in BY-RBM in the LV. The LV valve geometrical configuration, where the mitral and aortic valves are quite close together is probably the main reason for the OT-RBM and BY-RBM similarities in the LV. Over all the studied regions OT-RBM fibers were closer to BY-RBM ones, comparing to ST-RBM results. 

The different RBMs were indirectly compared running electrophysiological simulations involving the outflow tracts guided by the different RBM-based fiber configurations. Results of this comparison can be seen in  Figure~\ref{fig: Sensitivity}. One of the major contributions of the OT-RBM is that it greatly improved OTVA simulations when compared to clinical observations. The simulated longitudinal/perpendicular clinical ratios of the isochrones (Table~\ref{Table} and first column of Figure~\ref{fig: Sensitivity}) were in agreement with measurements made by clinicians during the intervention \cite{Acosta2015ImpactArrhythmias}. According to this study, clinical ratios for RVOT isochrones are 1.9$\pm$1.2 (n=14) and 0.79$\pm$0.4 (n=23) for LVOT isochrones. The OT-RBM isochrone clinical ratio for RVOT cases in our simulation study (1.86$\pm$0.22; see Table~\ref{Table}) was very close to the value measured by electrophysiologists. This agreement was possible due to the new distribution of fibers, especially in the RV and OTs, following histological observations. Other RBMs use simpler rules for generating fibers in these regions, obtaining unrealistic fiber configuration and, therefore, incorrect electric wave propagation and resulting isochrones (a ratio of 0.57$\pm$0.16). 

The influence of the fiber orientation in electrophysiological simulations was further analyzed in a sensitivity analysis (Figure~\ref{fig: Sensitivity}). Isochrones obtained using the OT-RBM (90$^\circ$) were the only ones showing good agreement with clinical observations from Electro-Anatomical Mapping (EAM) \cite{Acosta2015ImpactArrhythmias,Herczku2012MappingActivation.}. When the origin of the arrhythmia is in the RVOT (top row of Figure~\ref{fig: Sensitivity}), the longest axis is in the longitudinal direction of the RVOT, which corresponds to axis ratios larger than 1.5.  By contrast, the remaining fiber configurations showed an isochronal principal direction opposite to clinical observations, as observed in Figure~\ref{fig: Sensitivity}. Obtaining simulated isochrones similar to clinical ones by only following histological observations, while being blind to any other patient-specific electrophysiological data, is a strong feature of the proposed method.

Finally, we also compared the RBMs with fibers obtained from DT-MRI data. An ex-vivo human heart geometry from Johns Hopkins database was used for the comparison. Unfortunately, fiber orientation data was only available below a certain basal plane, substantially below the outflow tracts and the valves, preventing a detailed evaluation of the OT-RBM. Despite this limitation, the comparison showed relatively small fiber angle differences in the LV ($\sim$30$^\circ$; Figure~\ref{fig: Diff},d), except close to the apex, which is a complex region for all RBMs when compared to DT-MRI data. The small fiber angle differences between OT-RBM, BY-RBM and DT-MRI data suggests that they will have a similar behavior in simulations involving the LV~\cite{Arevalo2016,Bayer2012AModels,Deng2015AccuracyMRI}. In the RV, differences to DT-MRI fibers obtained with the OT-RBM are slightly lower than the ones obtained with other RBMs, especially in the septum, which has high relevance in the OTVAs. It is important to highlight that the OT-RBM is not designed to strictly follow DT-MRI data (neither in the LV nor in the RV), but guided by histology-derived information.
As can be seen in histological pictures shown in Figure~\ref{fig: Hist}, fibers in the RV endocardium and the OTs present a different orientation compared to those obtained by DT-MRI. The OT-RBM has been parameterized to match as much as possible fiber orientation observed in histological data and based on cardiac anatomist observations; by design, RV and OT fiber configuration from our RBM will necessarily differ from DT-MRI. Therefore, we cannot consider fiber angle differences between them as a conclusive index to assess the accuracy of the OT-RBM

Another major outcome of our study is the improved realism of OTVA electrophysiological simulations when guided by OT-RBM fiber configurations since it could support electrophysiologists in the prediction of the SOO in OTVA patients. Frequently, LVOT vs RVOT prediction of SOO in OTVAs is a challenging task for electrophysiologists, especially in difficult cases where the LVOT and RVOT geometries are spatially close and their electrophysiological signals from the electrocardiogram cannot be distinguished. Several works~\cite{Acosta2015ImpactArrhythmias,Herczku2012MappingActivation.,Yoshida2014AIndex,Zhang2017ValueRatio} proposed several indices to predict the SOO of ectopic focus and then reduce the time of and improve the success rate of interventions. Using OT-RBM based electrophysiological simulations on patient-specific geometries, we have been able to differentiate the SOO by measuring the longitudinal/perpendicular clinical ratio of the 10 ms isochrones. According to Table~\ref{Table} there are statistical significant differences between RVOT and LVOT (LCC or RCC) origin isochrone ratios. Isochronal areas can also be used to predict the SOO in simulations (larger in LVOT cases) but cannot be consistently be measured in clinical routine. The previous findings show the potential of our modelling pipeline to be used as an additional pre-operative source of information to the clinician to predict the site of origin of ectopic foci.

 One of the main limitations of the method is the absence of quantitative 3D fiber information from imaging data in the RV and in the OTs. We developed our method being guided by findings in histological studies from 2D pictures of ex-vivo specimens and following recommendations from anatomists involved in this work. It would be desirable to compare our results with data obtained from advanced imaging techniques such as micro-CT or X-ray phase-contrast imaging, especially from the RV and OTs, which is rarely available in the literature. 
In this work we also compared our results with ex-vivo cardiac DT-MRI. This technique provides information related to myofiber orientation, but it cannot be considered as ground-truth since measurements are quite noisy, show poor signal-to-noise ratio and change under gradient calibration and temperature variation~\cite{Teh2016ResolvingImaging}. In addition, smoothing and interpolation techniques required to reconstruct tensor data from cardiac DT-MRI do not allow capturing myofiber orientation changes in very heterogeneous areas. Unlike in the LV, where DT-MRI may provide a good approximation of myofibers in most areas, this is particularly critical in thin structures of the heart such as the outflow tracts and the RV. 
Another limitation of the developed RBM is the smoothness of the resulting fiber orientation distributions, which is inherent to RBM. Unlike the variability found in histological data, RBM tend to generate idealistic fiber orientation, in part due to the use of too smooth biventricular geometries (e.g. without trabeculae). 
Finally, the definition of the longitudinal direction in some points of the boundary surfaces can affect the fiber orientation. Due to the definition of the boundary, the gradient in these points is zero, thus obtaining an incorrect fiber orientation (an example can be seen in Figure~\ref{fig: Diff}, where apical regions have a slightly higher angle difference in all cases). However, this problem can be minimized by reducing the number of points taken as boundaries and applying interpolation with the closest neighbors.

\section{Conclusion}

We have presented a RBM that includes fiber information specific to the RV, interventricular septum and both OT of the ventricles, to replicate histological observations. The proposed model facilitates the use of electrophysiological simulations in applications when these regions are important such as in OTVA patients, which was not possible before. The method has been indirectly validated comparing the results of electrophysiological simulations using fiber orientation provided by the proposed RBM with clinical observations in OTVA patients. These obtained results showed a very good agreement with clinical observations, demonstrating the potential of the developed modelling pipeline as a tool for the prediction of the site of origin of the ectopic focus in this type of arrhythmias. Further work will be focused on the application of this pipeline to a larger clinical database and in the adaptation of the developed RBM to high-resolution fiber information provided by X-ray phase-contrast imaging data.

\section{Bibliography}

%\bibliographystyle{elsarticle-harv}

%\bibliography{Mendeley} 
%\clearpage
\appendix
\onecolumn
\section{Quantitative analysis of fiber differences}\label{app.p}
 \begin{figure}[h!]
\centering
\includegraphics[scale=0.5]{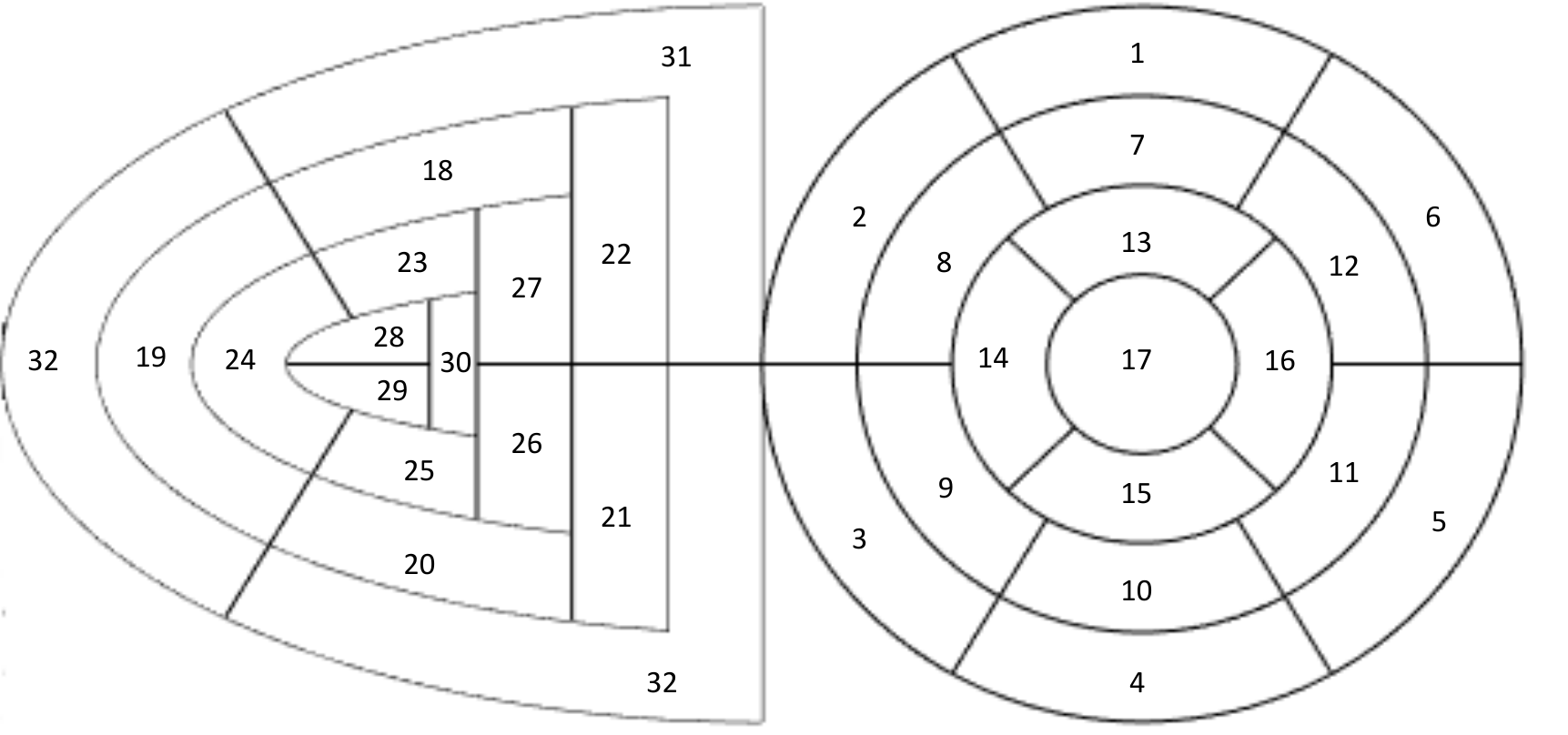}
\caption{2D plot of the biventricular geometry divided in 32 regions}
\label{fig: Aha}
\captionof{table}{Quantitative comparison of the angle differences presented in Figure~\ref{fig: Diff} (b-f). Mean angle difference and standard deviation are presented for each region described in Figure~\ref{fig: Diff} (a)}. 
\begin{tabular}{cccccc}
\multicolumn{6}{c}{\textbf{angle  $\pm$  sd    ($^\circ$)}} \\ \hline
\multicolumn{1}{c|}{\textbf{Region}} & \textbf{ST-RBM vs OT-RBM} & \textbf{BY-RBM vs OT-RBM} & \textbf{DTI vs OT-RBM} & \textbf{DTI vs ST-RBM} & \textbf{DTI vs BY-RBM} \\ \hline
\multicolumn{1}{c|}{\textbf{1}} & 23,0 $\pm$ 14,0 & 19,8 $\pm$ 12,9 & 26,6 $\pm$ 16,0 & 23,0 $\pm$ 15,0 & 16,3 $\pm$ 12,8 \\
\multicolumn{1}{c|}{\textbf{2}} & 26,4 $\pm$ 18,6 & 18,3 $\pm$ 19,5 & 25,5 $\pm$ 14,8 & 32,9 $\pm$ 17,3 & 25,3 $\pm$ 15,4 \\
\multicolumn{1}{c|}{\textbf{3}} & 31,6 $\pm$ 19,8 & 23,0 $\pm$ 24,5 & 22,0 $\pm$ 19,2 & 31,5 $\pm$ 17,9 & 17,0 $\pm$ 15,1 \\
\multicolumn{1}{c|}{\textbf{4}} & 21,5 $\pm$ 12,2 & 15,3 $\pm$ 14,3 & 29,1 $\pm$ 13,4 & 25,8 $\pm$ 13,3 & 20,6 $\pm$ 14,0 \\
\multicolumn{1}{c|}{\textbf{5}} & 28,6 $\pm$ 10,8 & 17,3 $\pm$ 14,4 & 29,5 $\pm$ 14,8 & 26,2 $\pm$ 10,8 & 19,6 $\pm$ 11,7 \\
\multicolumn{1}{c|}{\textbf{6}} & 28,8 $\pm$ 11,1 & 18,8 $\pm$ 14,1 & 25,1 $\pm$ 12,2 & 20,0 $\pm$ 11,2 & 15,0 $\pm$ 10,9 \\
\multicolumn{1}{c|}{\textbf{7}} & 39,0 $\pm$ 10,0 & 18,5 $\pm$ 10,0 & 33,1 $\pm$ 16,5 & 33,9 $\pm$ 10,0 & 17,6 $\pm$ 14,3 \\
\multicolumn{1}{c|}{\textbf{8}} & 43,3 $\pm$ 16,8 & 23,4 $\pm$ 18,5 & 40,8 $\pm$ 17,3 & 37,5 $\pm$ 12,6 & 26,6 $\pm$ 14,6 \\
\multicolumn{1}{c|}{\textbf{9}} & 38,8 $\pm$ 13,6 & 12,5 $\pm$ 12,1 & 32,7 $\pm$ 21,1 & 44,2 $\pm$ 13,0 & 28,5 $\pm$ 22,5 \\
\multicolumn{1}{c|}{\textbf{10}} & 39,5 $\pm$ 12,7 & 11,0 $\pm$ 7,1 & 22,9 $\pm$ 15,2 & 41,4 $\pm$ 8,3 & 17,9 $\pm$ 17,1 \\
\multicolumn{1}{c|}{\textbf{11}} & 48,9 $\pm$ 12,2 & 11,1 $\pm$ 7,8 & 27,2 $\pm$ 15,0 & 46,2 $\pm$ 9,1 & 18,7 $\pm$ 12,3 \\
\multicolumn{1}{c|}{\textbf{12}} & 46,0 $\pm$ 11,3 & 14,1 $\pm$ 10,2 & 25,2 $\pm$ 12,9 & 43,9 $\pm$ 10,8 & 14,2 $\pm$ 10,5 \\
\multicolumn{1}{c|}{\textbf{13}} & 57,6 $\pm$ 11,1 & 16,0 $\pm$ 9,5 & 44,3 $\pm$ 22,1 & 57,6 $\pm$ 13,8 & 30,3 $\pm$ 16,9 \\
\multicolumn{1}{c|}{\textbf{14}} & 58,7 $\pm$ 14,7 & 27,7 $\pm$ 18,3 & 36,4 $\pm$ 20,2 & 55,4 $\pm$ 12,1 & 28,2 $\pm$ 20,1 \\
\multicolumn{1}{c|}{\textbf{15}} & 59,9 $\pm$ 11,9 & 14,4 $\pm$ 10,8 & 30,5 $\pm$ 20,6 & 56,7 $\pm$ 7,8 & 23,1 $\pm$ 17,4 \\
\multicolumn{1}{c|}{\textbf{16}} & 69,6 $\pm$ 10,4 & 13,3 $\pm$ 10,9 & 46,7 $\pm$ 21,4 & 66,5 $\pm$ 9,0 & 36,8 $\pm$ 16,1 \\
\multicolumn{1}{c|}{\textbf{17}} & 72,5 $\pm$ 10,1 & 28,7 $\pm$ 24,0 & 33,5 $\pm$ 20,8 & 62,6 $\pm$ 10,4 & 39,1 $\pm$ 22,8 \\
\multicolumn{1}{c|}{\textbf{18}} & 15,9 $\pm$ 8,1 & 12,5 $\pm$ 8,4 & 13,4 $\pm$ 10,6 & 16,0 $\pm$ 8,3 & 14,6 $\pm$ 8,8 \\
\multicolumn{1}{c|}{\textbf{19}} & 15,1 $\pm$ 5,5 & 15,1 $\pm$ 11,7 & 21,6 $\pm$ 14,0 & 20,0 $\pm$ 11,6 & 16,9 $\pm$ 13,2 \\
\multicolumn{1}{c|}{\textbf{20}} & 23,6 $\pm$ 7,4 & 33,0 $\pm$ 13,3 & 19,7 $\pm$ 18,3 & 29,5 $\pm$ 14,6 & 34,5 $\pm$ 16,5 \\
\multicolumn{1}{c|}{\textbf{21}} & 41,2 $\pm$ 16,0 & 43,4 $\pm$ 22,9 & 36,4 $\pm$ 25,3 & 43,6 $\pm$ 17,9 & 42,7 $\pm$ 23,7 \\
\multicolumn{1}{c|}{\textbf{22}} & 45,7 $\pm$ 25,6 & 44,6 $\pm$ 29,2 & 27,4 $\pm$ 17,2 & 28,8 $\pm$ 13,8 & 23,3 $\pm$ 14,0 \\
\multicolumn{1}{c|}{\textbf{23}} & 20,4 $\pm$ 6,5 & 13,0 $\pm$ 6,4 & 24,8 $\pm$ 13,8 & 21,6 $\pm$ 13,6 & 24,9 $\pm$ 16,3 \\
\multicolumn{1}{c|}{\textbf{24}} & 23,4 $\pm$ 5,3 & 13,0 $\pm$ 9,5 & 25,1 $\pm$ 16,3 & 26,9 $\pm$ 11,6 & 21,0 $\pm$ 13,6 \\
\multicolumn{1}{c|}{\textbf{25}} & 29,0 $\pm$ 8,2 & 34,8 $\pm$ 15,2 & 21,2 $\pm$ 11,5 & 34,6 $\pm$ 7,7 & 35,3 $\pm$ 12,3 \\
\multicolumn{1}{c|}{\textbf{26}} & 52,9 $\pm$ 17,2 & 51,9 $\pm$ 26,0 & 38,3 $\pm$ 25,0 & 47,0 $\pm$ 13,2 & 47,3 $\pm$ 21,3 \\
\multicolumn{1}{c|}{\textbf{27}} & 45,3 $\pm$ 26,5 & 39,4 $\pm$ 28,5 & 40,4 $\pm$ 17,7 & 28,7 $\pm$ 15,9 & 34,0 $\pm$ 21,2 \\
\multicolumn{1}{c|}{\textbf{28}} & 39,1 $\pm$ 15,1 & 22,2 $\pm$ 17,1 & 37,3 $\pm$ 20,5 & 34,7 $\pm$ 13,9 & 28,0 $\pm$ 14,1 \\
\multicolumn{1}{c|}{\textbf{29}} & 52,0 $\pm$ 18,9 & 42,1 $\pm$ 27,6 & 48,0 $\pm$ 25,0 & 35,1 $\pm$ 7,4 & 25,5 $\pm$ 11,3 \\
\multicolumn{1}{c|}{\textbf{30}} & 56,0 $\pm$ 21,3 & 40,8 $\pm$ 26,0 & 47,6 $\pm$ 22,9 & 42,0 $\pm$ 19,3 & 48,3 $\pm$ 21,4 \\
\multicolumn{1}{c|}{\textbf{31}} & 31,9 $\pm$ 20,6 & 31,5 $\pm$ 21,2 & 23,4 $\pm$ 19,7 & 28,0 $\pm$ 20,8 & 24,5 $\pm$ 19,8 \\
\multicolumn{1}{c|}{\textbf{32}} & 28,7 $\pm$ 19,2 & 30,7 $\pm$ 20,0 & 18,8 $\pm$ 15,7 & 26,0 $\pm$ 17,9 & 23,4 $\pm$ 17,9
\end{tabular}
\label{tab2}
\end{figure}

\end{document}